# Creation of Coherent Superposition States in Inhomogeneously Broadened Media with Relaxation

N. Sandor, J.S. Bakos, Zs. Sörlei and G.P. Djotyan*

*Research Institute for Particle & Nuclear Physics, Budapest, POBox 49, Hungary;*

*Corresponding author: djotjan@rmki.kfki.hu

Abstract. We propose and analyze a scheme for "on demand" creation of coherent superposition of meta-stable states in a tripod-structured atom using frequency-chirped laser pulses. Negligible excitation of the atoms during the creation of the superposition states is a priority in our consideration. The underlying physics of the scheme is explained using the formalism of adiabatic states. By numerically solving master equation for the density matrix operator, we analyze the influence of the spontaneous decay and transverse relaxation on the efficiency of the creation of superposition states. We show that the proposed scheme is robust against small-to-medium variations of the parameters of the laser pulses. We provide a detailed analysis of the effect of the inhomogeneous (Doppler-) broadening on the efficiency of the coherence creation and show that the proposed scheme may be equally efficient in both homogeneously and Doppler-broadened media.
32.80 Qk, 42.50 Gy

## 1. Introduction

Creation of coherent superposition of quantum states attracted a great amount of attention in the last decade. The reason is a rich variety of possibilities for practical applications of coherent superposition states in different fields of the optical science and technology. Preparation of the atoms or molecules in coherent superposition states results in extreme changes in the optical (refractive and absorption) properties of a medium composed of such coherently prepared atoms or molecules. These changes may lead to dramatic variations of the propagation characteristics of laser pulses in such media due to the effect of electromagnetically induced transparency (see review papers [1-3] and References therein). Other effects and applications relating to the coherently prepared media include inversionless amplification of laser radiation [4], enhancement of the efficiency of nonlinear optical processes [5-7], quantum computing [8], optical information processing using quantum states [9-16] and many others.

There is a large variety of techniques for creation of coherent superposition of atomic states. In a simplest case of two-level model atom, for example, a coherent superposition of the two states may be created using a resonant laser pulse with the area of the Rabi frequency (integral of the Rabi frequency over time) not equal to an integer number of $\pi$. The value of the induced coherence however is sensitive even to relatively small variations of this area and to the resonance conditions between the interacting laser field and atomic transition [17].

Other, more robust schemes for creation of coherent superposition states were proposed in the last decade in atomic and molecular systems with more complicated structure of working levels. The majority of these schemes are based on the adiabatic following method, which is robust against small-to-medium variations of the parameters of the laser fields [18-29]. The mostly used schemes of the adiabatic following include stimulated Raman adiabatic passage (STIRAP) [18-26], Stark chirped rapid-adiabatic passage (SCRAP) [27-29] and schemes involving frequency modulated (chirped) laser radiation [9, 30-38].

The condition of two-photon resonance is crucial for the STIRAP scheme to be effective. Being sensitive to this condition, the STIRAP – based schemes allow negligible excitation of the atomic system. The SCRAP scheme is not too sensitive to the two-photon resonance condition but is accompanied by temporary excitation of the atom. The schemes based on the frequency chirped (FC) laser pulses are free of the drawbacks of the both mentioned schemes.

Due to the frequency chirp they are not sensitive to the resonance conditions (including the two-photon resonance condition) and allow transition of the populations between the metastable states of a multilevel atom without considerable excitation of the atom [31-33]. These features of the schemes based on the FC pulses are especially important for the coherent manipulation schemes of inner or (and) translational states of quantum systems, where excitation is prohibited to avoid the de-coherence effects caused by the spontaneous decay of the excited states. Such elimination of the excitation in multilevel atoms is achieved through destructive interference of different quantum passes to the excited state.

In [31], a novel scheme of manipulation of atomic populations was proposed and a complete population transfer between meta-stable states of the Λ-type atom was demonstrated without considerable excitation of the atom by using a single FC laser pulse. This scheme may be considered as a supplementary one to the STIPAP scheme. While the condition of Raman resonance is crucial for the STIRAP-based schemes to be effective, it *must be violated* in the scheme with chirped laser pulse of Ref.[31] to avoid excitation of the atom. This is a major difference of this scheme compared with that of the STIRAP. As our analysis shows, in the scheme of Ref.[31] the initial mixture of the bright and dark superposition states when the atom is optically pumped into one of the ground states of the Λ-like atom evolves into a completely dark state at the central part of the laser pulse. At the end of the interaction, the same mixture of the dark and bright states as in the initial superposition is created but with additional phase difference of $\pi$. It results in complete transfer of the atomic population from the initially populated ground state to the other initially empty ground state without considerable population of the excited state. A similar scheme with a single FC pulse but in a tripod-type atom was used in [32] to transfer the atomic populations between the three metastable states and to create a coherent superposition of these states without considerable excitation of the atom. Note that while the scheme proposed in Ref.[33] using STIRAP with one of the pulses having chirped frequency allows creation of an arbitrary coherence through variation of the speed of the chirp, it contains a non-adiabatic coupling between the dressed states and hence, is not purely adiabatic and a robust one.

From the point of view of practical applications of schemes for coherent manipulation of atomic or molecular quantum states, it is important to investigate their applicability in real media with inhomogeneously broadened transition lines typical for atomic gases or atoms in solid-state environment. Another important question to be addressed is the effect of the longitudinal and transverse relaxations on the efficiency of the schemes. Note that in the case of STIRAP the effect of dephasing was investigated in Refs. [39,40] and recently, in Ref.[41]. The influence of the dephasing and inhomogeneous line broadening due to solvent fluctuations on the efficiency of STIPAP in solute molecules was studied in Ref. [42].

In this paper, we generalize the scheme proposed in Refs. [31, 32] by including three FC laser pulses (instead of a single one) and thus, increasing the potential to control atomic populations and created coherences. No time delay between the pulses is needed for the scheme under consideration that reveals an additional dissimilarity with the STIRAP-based schemes. Since the proposed scheme includes FC laser pulses and doesn't require strict resonance conditions, it seems to be perspective for applications in inhomogeneously broadened media. To explore the potential possibilities of the scheme, we investigate creation of coherent superposition of meta-stable (ground) states in a tripod- structured model atom with Doppler- broadened transition lines.

In the proposed scheme, three FC laser pulses with two of them in Raman resonance and the third one out of Raman resonance perform coherent manipulation of the atomic populations and coherences. Similarly to the case of a single FC laser pulse (see [31, 32]) the proposed scheme allows suppressing of the excitation of the atom to minimize the decoherence effects caused by the spontaneous decay of the excited states.

We investigate in details the effect of Doppler-broadening of transition lines on the efficiency of coherence creation between ground atomic states by considering an atomic gas of tripod-structured atoms at different temperatures. The influence of the relaxation processes in the atomic ensemble on the efficiency of the coherence creation is analyzed as well to adequately model a real experimental situation.

We show that while the proposed scheme is robust to variations of the parameters of the laser field out of Raman resonance, the value of the created coherence may be controlled by variation of amplitudes (Rabi frequencies) of the two laser pulses in Raman resonance. In the case when these two laser pulses originate from a same laser source, the induced coherence is immune to phase and amplitude fluctuations of these two fields.

The remainder of this paper is organized as follows. In Section 2, we present the mathematical formalism describing the interaction of FC laser pulses with the tripod-structured atom based on the master equations for the density matrix operator taking into account longitudinal and transverse relaxation processes. In Section 3, creation of the coherent superposition states is discussed using dressed states analysis. The results of the numerical analysis are presented in Section 4. The effect of Doppler-broadening of the transition lines in an atomic gas of the tripod-structured atoms is analyzed in Section 5. In Section 6, the influence of longitudinal and transverse relaxation processes is discussed on the efficiency of the coherence creation. The robustness of the proposed scheme against variations of parameters of the acting laser pulses is analysed in Section 7, and the obtained results are summarized in Section 8.

## 2. The Mathematical Formalism

We consider interaction of three laser pulses with a model atom having tripod-like structure of working levels, see Fig.1. Each laser field is acting on corresponding allowed electric-dipole transition in the atom (between a ground state $|n\rangle$, ($n = 1, 2, 3$) and the excited state $|0\rangle$) according to selection of the carrier frequencies or polarizations of the laser fields. Transitions between the ground states are forbidden in the electric-dipole approximation.

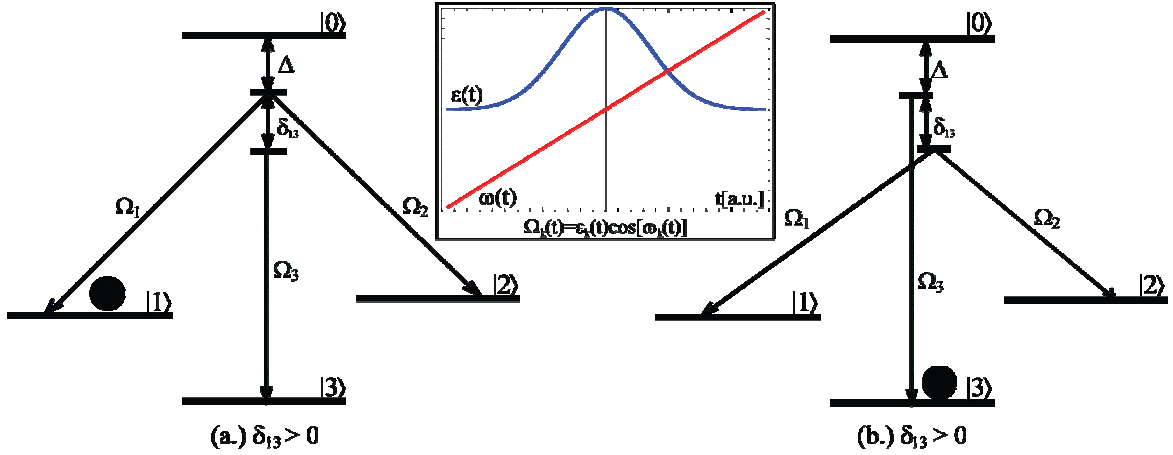

Fig.1 Level structure of the tripod-like atom and the interacting FC laser pulses for two cases of positive (a) and negative (b) Raman detuning and two different initial conditions for the atomic population. The inset – a Gaussian laser pulse with positive linear chirp.

The electric field strength of the laser field has the form $E(t) = \sum_{i=1}^{3} E_i(t) \cos\left( \int_{-\infty}^{t} \omega_i(t') dt' \right)$. A same linear variation in time (linear frequency chirp) of the carrier frequencies of all the interacting laser pulses is assumed in this paper: $\omega_k(t) = \omega_k^{(0)} + \beta t$, where $\omega_k^{(0)}$ is the central carrier frequency of the $k^{th}$ field and $\beta$ is the speed of the chirp. The central frequency detuning $\delta_k$ from the resonance atomic frequency $\omega_{0k}$ is $\delta_k = \omega_{0k} - \omega_k^{(0)}$, $k \in \{1,2,3\}$. Since we assume Raman resonance between the fields $E_1$ and $E_2$, we have equal single-photon detuning for these waves: $\delta_1 = \delta_2 = \Delta$, see Fig.1.

Since we are interested in the effect of the relaxation processes on the dynamics of the atomic populations and coherences, we begin with analysis of the master equation for the density operator $\hat{\rho}$ of the system:

$$i\hbar \partial_t \hat{\rho} = \left[ \hat{H}, \hat{\rho} \right] + \hat{R}, \tag{1}$$

where the interaction Hamiltonian in the rotating wave approximation has the following form:

$$\hat{H} = \hbar \begin{pmatrix} 0 & e^{i\Delta t}\Omega_1 & e^{i\Delta t}\Omega_2 & e^{i\delta_3 t}\Omega_3 \\ e^{-i\Delta t}\Omega_1^* & 0 & 0 & 0 \\ e^{-i\Delta t}\Omega_2^* & 0 & 0 & 0 \\ e^{-i\delta_3 t}\Omega_3^* & 0 & 0 & 0 \end{pmatrix} \quad (2)$$

and the relaxation operator is:

$$\hat{R} = i\hbar \begin{pmatrix} -\Gamma\rho_{00} & (-\gamma_{01}+\Gamma/2)\rho_{01} & (-\gamma_{02}+\Gamma/2)\rho_{02} & (-\gamma_{03}+\Gamma/2)\rho_{03} \\ (-\gamma_{01}+\Gamma/2)\rho_{01}^* & \Gamma_1\rho_{00} & -\gamma_{12}\rho_{12} & -\gamma_{13}\rho_{13} \\ (-\gamma_{02}+\Gamma/2)\rho_{02}^* & -\gamma_{12}\rho_{12}^* & \Gamma_2\rho_{00} & -\gamma_{23}\rho_{23} \\ (-\gamma_{03}+\Gamma/2)\rho_{03}^* & -\gamma_{13}\rho_{13}^* & -\gamma_{23}\rho_{23}^* & \Gamma_3\rho_{00} \end{pmatrix} \quad (3)$$

The time-dependent Rabi frequencies are $\Omega_j = \dfrac{E_j(t) d_{0j} \cdot e^{i\beta t^2}}{\hbar} = W_j f(t) \cdot e^{i\beta t^2}$, with (in general, different) complex amplitudes $W_j$, ($j = 1, 2, 3$). The envelope function $f(t)$ is supposed to be the same for the three laser pulses. $d_{0j}$ is the dipole moment matrix element for transition between the ground state $|j\rangle$ and the excited state $|0\rangle$. We assume that the laser pulses have a same Gaussian shape with duration $\tau_p = \tau_L/(\sqrt{2\ln 2})$, where $\tau_L$ is the full width at half maximum (FWHM) of the pulse (intensity) envelope: $f(t) = \exp\left[-(t/\tau_p)^2\right]$. $\Gamma = \Gamma_{01} + \Gamma_{02} + \Gamma_{03}$ is the spontaneous decay rate of the excited state $|0\rangle$ with $\Gamma_{0j}$ being spontaneous relaxation rate to the ground state $|j\rangle$; $\gamma_{0j}$ is the dephasing rate for optical coherence between the states $|0\rangle$ and $|j\rangle$, ($j = 1, 2, 3$); and $\gamma_{12}, \gamma_{13}, \gamma_{23}$ are dephasing rates of coherences between the ground states. The following phase transformation of the density matrix operator: $\rho' = T^{-1} \cdot \rho \cdot T$ with

$$T = \begin{pmatrix} 1 & 0 & 0 & 0 \\ 0 & e^{-i\Delta} & 0 & 0 \\ 0 & 0 & e^{-i\Delta} & 0 \\ 0 & 0 & 0 & e^{-i\delta_3} \end{pmatrix}$$

allows us to eliminate the exponential phase factors in the right-side of the Eq.(1) and in the Hamiltonian in Eq.(2):

$$H' = T^{-1}HT + i\hbar\left(\partial_t T^{-1}\right)T = \begin{pmatrix} 0 & \Omega_1 & \Omega_2 & \Omega_3 \\ \Omega_1^* & -\Delta & 0 & 0 \\ \Omega_2^* & 0 & -\Delta & 0 \\ \Omega_3^* & 0 & 0 & -\delta_3 \end{pmatrix}.$$

(4)

resulting in the following set of equations for the density matrix elements:

$$\partial_\tau \rho_{00} = -i\left(\sum_{i=1}^{3}\tilde{\Omega}_i\rho_{0i}^* - \tilde{\Omega}_i^*\rho_{0i}\right) - \tilde{\Gamma}\rho_{00}$$

$$\partial_\tau \rho_{kk} = i\left(\tilde{\Omega}_k\rho_{01}^* - \tilde{\Omega}_k^*\rho_{01}\right) + \tilde{\Gamma}_k\rho_{00}, \text{ where } k \in \{1,2,3\}$$

$$\partial_\tau \rho_{01} = -i\left[\tilde{\Omega}_2\rho_{12}^* + \tilde{\Omega}_3\rho_{13}^* + \tilde{\Omega}_1(\rho_{11}-\rho_{00}) + \tilde{\Delta}\rho_{01}\right] - \left(\tilde{\Gamma}/2 + \tilde{\gamma}_{01}\right)\rho_{01}$$

$$\partial_\tau \rho_{02} = -i\left[\tilde{\Omega}_1\rho_{12} + \tilde{\Omega}_3\rho_{23}^* + \tilde{\Omega}_2(\rho_{22}-\rho_{00}) + \tilde{\Delta}\rho_{02}\right] - \left(\tilde{\Gamma}/2 + \tilde{\gamma}_{02}\right)\rho_{02} \qquad (5)$$

$$\partial_\tau \rho_{03} = -i\left[\tilde{\Omega}_1\rho_{13} + \tilde{\Omega}_2\rho_{23} + \tilde{\Omega}_3(\rho_{33}-\rho_{00}) + \tilde{\delta}_3\rho_{03}\right] - \left(\tilde{\Gamma}/2 + \tilde{\gamma}_{03}\right)\rho_{03}$$

$$\partial_\tau \rho_{12} = i\left[\tilde{\Omega}_2\rho_{01}^* - \tilde{\Omega}_1^*\rho_{02}\right] - \tilde{\gamma}_{12}\rho_{12}$$

$$\partial_\tau \rho_{13} = i\left[\tilde{\Omega}_3\rho_{01}^* - \tilde{\Omega}_1^*\rho_{03} + \tilde{\delta}_{13}\rho_{13}\right] - \tilde{\gamma}_{13}\rho_{13}$$

$$\partial_\tau \rho_{23} = i\left[\tilde{\Omega}_3\rho_{02}^* - \tilde{\Omega}_2^*\rho_{03} + \tilde{\delta}_{13}\rho_{23}\right] - \tilde{\gamma}_{23}\rho_{23},$$

where a dimensionless time variable $\tau = \dfrac{t}{\tau_p}$ is introduced and all the parameters are scaled with $\tau_p$ (or $1/\tau_p$): $\tilde{\Gamma}_j = \Gamma_j \tau_p$, $\tilde{\gamma}_{kl} = \gamma_{kl} \tau_p$, $\tilde{\Delta} = \Delta \tau_p$. $\tilde{\delta}_{13} = (\Delta - \delta_3)\tau_p$ is Raman detuning between the fields with Rabi frequencies $\Omega_{1,2}$ and $\Omega_3$, $\tilde{\Omega}_k(\tau) = \tilde{W}_k\, e^{-\tau^2} e^{\tilde{\beta}\tau^2}$ with $\tilde{W}_k = W_k \tau_p$; $\tilde{\beta} = \beta \tau_p^2$.

Before presenting and analyzing the results of the numerical simulation of Eqs.(5), and to gain an insight into the underlying physics of the processes in the atomic system, let us discuss the behaviour of the system in the case when the relaxation processes may be neglected assuming duration of the pulses much shorter than the relaxations times scales.

At this point it is convenient to perform a transformation from the bare state representation basis for the atomic state vector to the basis of "dark"-"bright" superposition states and perform the dressed states analysis in that basis.

**2.1. Transformation to the "dark"-"bright" superposition states basis**

For the analysis provided below, it is convenient to introduce new basis functions consisting of the following superpositions of the bare atomic states of the tripod-like atom:

$$\left|db_1\right\rangle = \dfrac{W_2^*\left|1\right\rangle - W_1^*\left|2\right\rangle}{\sqrt{\left|W_1\right|^2 + \left|W_2\right|^2}}; \quad \left|db_2\right\rangle = \dfrac{W_1\left|1\right\rangle + W_2\left|2\right\rangle}{\sqrt{\left|W_1\right|^2 + \left|W_2\right|^2}}; \tag{6}$$

$$\left|db_3\right\rangle = \dfrac{W_3\left|3\right\rangle}{\left|W_3\right|}; \quad |db_4\rangle \equiv |0\rangle.$$

In the new basis, the Hamiltonian of Eq.(4) can be written as:

$$\hat{H}_{db} = \hbar \begin{pmatrix} 0 & 0 & 0 & 0 \\ 0 & 0 & 0 & f(t)\sqrt{\left|W_1\right|^2 + \left|W_2\right|^2} \\ 0 & 0 & \delta_{13} & f(t)\left|W_3\right| \\ 0 & f(t)\sqrt{\left|W_1\right|^2 + \left|W_2\right|^2} & f(t)\left|W_3\right| & 2t\beta + \Delta \end{pmatrix}. \tag{7}$$

## 3. Creation of Coherent Superposition States: Dressed States Analysis

The state vector of the atom in the dark-bright basis can be represented on the basis of the eigenfunctions of the interaction Hamiltonian $\hat{H}_{db}$ as:

$$|\psi(t)\rangle = \sum_k r_k(t)\vec{b}_k(t) \cdot e^{\int_{-\infty}^{t} \lambda_k(t_1)dt_1}, \qquad (8)$$

with inital condition $|\psi(-\infty)\rangle = \sum_k r_k(-\infty)\vec{b}_k(-\infty)$, where $\lambda_k$ and $\vec{b}_k$ are the eigen-values and eigen-vectors of the Hamiltonian $\hat{H}_{db}$:

$$\hat{H}_{db}\vec{b}_k = \lambda_k \vec{b}_k. \qquad (9)$$

According to the adiabatic theorem, (see Refs.[43, 44]) in the adiabatic regime of interaction $r_k(t) \equiv r_k(-\infty)$, $k \in \{1,2,3,4\}$, which means that if the atom is in one (or superposition) of its eigen-states initially (before the interaction process, at $t \to -\infty$), it remains in the same eigen-state (or the superposition of the eigen-states) during the interaction process.

We obtain the following equation for the eigen-values (quasienergies) $\lambda = \lambda_k$ from Eqs. (7-9):

$$\lambda(\lambda^3 + \lambda^2(\delta_{13} - 2\beta t - \Delta) - \lambda\left[\left(|W_1|^2 + |W_2|^2 + |W_3|^2\right)f(t)^2 + \delta_{13}(2\beta t + \Delta)\right] \\ -(|W_1|^2 + |W_2|^2)\delta_{13}f(t)^2) = 0 \qquad (10)$$

and the following expressions for the corresponding dressed eigen-vectors $\vec{b}_j$:

$$\vec{b}_k = \left(0 \quad \frac{f(t)\sqrt{|W_1|^2 + |W_2|^2}}{\sqrt{N}} \quad \frac{\lambda_k(\lambda_k - [\Delta + 2\beta t]) - f(t)^2(|W_1|^2 + |W_2|^2)}{f(t)|W_3|\sqrt{N}} \quad \frac{\lambda_k}{\sqrt{N}}\right)^{(T)}, k \in \{2,3,4\}, \quad (11)$$

and $\vec{b}_1 = (1 \ 0 \ 0 \ 0)^{(T)}$, with the normalization factor

$$N = \sqrt{f(t)^2 \left(|W_1|^2 + |W_2|^2\right) + \left|\frac{\lambda_k \left(\lambda_k - [\Delta + 2\beta t]\right) - f(t)\left(|W_1|^2 + |W_2|^2\right)}{f(t)|W_3|}\right|^2 + \lambda_k^2}\ .$$

As it follows from Eq.(11), the following relations take place between the components of the dressed state vector $\vec{b}_k, k \in \{2,3,4\}$:

$$\frac{b_k^{(3)}}{b_k^{(2)}} = \frac{\lambda_k \left(\lambda_k - [\Delta + 2\beta t]\right) - f(t)\left(|W_1|^2 + |W_2|^2\right)}{f(t)^2 |W_3|\sqrt{|W_1|^2 + |W_2|^2}}, \quad \frac{b_k^{(4)}}{b_k^{(2)}} = \frac{\lambda_k}{f(t)\sqrt{|W_1|^2 + |W_2|^2}}. \tag{12}$$

The dynamics of the four eigen-values (quasienergies) is shown in Fig.2 for Gaussian shaped laser pulses with positive ($\beta>0$) linear chirp for two cases of Raman-detuning $\delta_{13}$: positive ($\delta_{13}>0$) and negative ($\delta_{13}<0$) one.

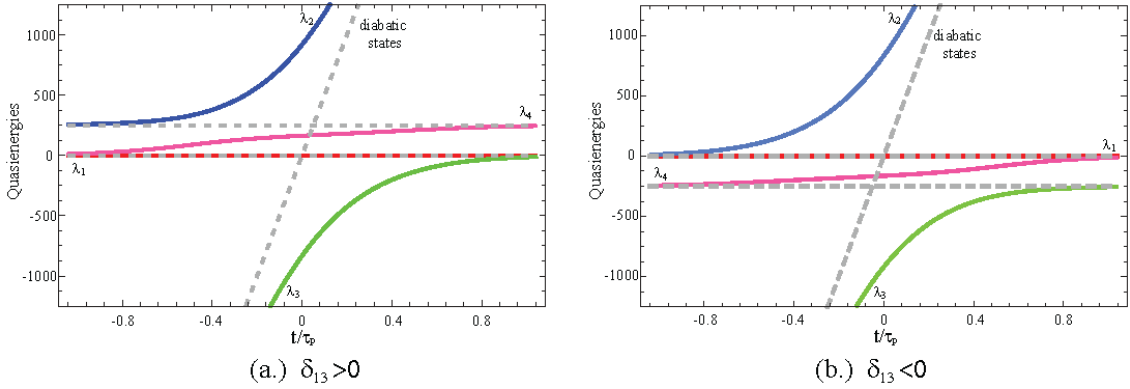

(a.) $\delta_{13}>0$          (b.) $\delta_{13}<0$

Fig. 2. Time dependence of the eigen-values (quasi-energies) of Hamiltonian (14) in case of positive (a) and negative (b) Raman detuning; $|\delta_{13}| \cdot \tau_p = 250$ in both cases. All parameters are normalized to the pulse duration $\tau_p$.

Below we consider two different cases of creation of coherent superposition states that differ by the initial preparation of the atom (see Fig.1). In the first case, before the interaction process (at $t \to -\infty$), all the atomic population is optically pumped into one of the two ground states connected by the pulses in Raman-resonance (see Fig.1a). In the second case, the atom is prepared in the ground state $|3\rangle$ initially, (see Fig.1b). We will refer to the cases of positive and negative Raman detuning supposing that the atom is prepared in the ground state $|1\rangle$ (positive Raman detuning), or in the state $|3\rangle$ (negative Raman detuning) with a same positive chirp ($\beta > 0$).

While in the first case of the positive Raman detuning ($\delta_{13} > 0$), at the positive chirp ($\beta > 0$) the pulses in Raman resonance ($\Omega_1$ and $\Omega_2$) reach the single-photon resonances with corresponding transitions earlier than the third pulse ($\Omega_3$) out of Raman resonance, in the second case of negative Raman detuning (and positive chirp), the resonances occur in reversed order. As it will be shown below, in both cases of the initial preparation of the atom only a negligible (and temporary) excitation of the atom takes place during the interaction.

As it can be seen in Fig. 2, in both cases of the positive and negative Raman detuning, the quasienergy $\lambda_1$ has zero value at the beginning and during all the interaction process: $\lambda_1 = 0$. This quasienergy corresponds to the dark superposition state $|db_1\rangle$, (see Eq.(6)). In the same time, there is another quasienergy, $\lambda_4$ (see Fig.2), which value, in contrast to the other two quasienergies $\lambda_2$ and $\lambda_3$, does not depend strongly on the Rabi frequencies and is restricted by the value of the Raman detuning $\delta_{13}$:

$$0 \leq |\lambda_4| \leq |\delta_{13}| \qquad (13)$$

Let us assume at this point that before the interaction process, all the atomic population is distributed between the ground states $|1\rangle$ and $|2\rangle$ and is prepared in the bright superposition $|db_2\rangle$, (see Eq.(6)). As it follows from Fig.2(a), in the case of positive Raman detuning only the quasienergy $\lambda_4$ coincides at the beginning of the interaction ($t \to -\infty$) with the energy of the (bare) state $|db_2\rangle$; and the corresponding adiabatic (dressed) state $\vec{b}_4$ coincides with the state $|db_2\rangle$. It means that the dressed state $\vec{b}_4$ is one that has to be identified with the assumed initial state vector of the atom in the absence of the laser field. At the end of the interaction, this quasienergy coincides with the energy of the bare state $|db_3\rangle$, which with a phase factor (equal to $W_3/|W_3|$) coincides with the bare ground state $|3\rangle$ in the initial (energy) representation. It means that the atomic population of the "bright" superposition of two ground states $|1\rangle$ and $|2\rangle$ is adiabatically transferred into the third ground state of the atom as a result of interaction with the three frequency chirped laser pulses. The same scenario will take place in the case of negative Raman detuning if initially the atom is localized in the state $|3\rangle$, as in the second scheme of coherence creation (see Fig.2b).

In the case of positive Raman detuning, the atom is prepared in the atomic bare state $|1\rangle$, which is a superposition of the dark state $|db_1\rangle$ (that remains unchanged during the interaction), and the bright state $|db_2\rangle$, which evolves to ground state $|3\rangle$. Consequently, the atom ends up in a superposition of the dark state $|db_1\rangle$ and the state $|3\rangle$.

In order to provide coherent population transfer, it is important to avoid excitation of the atom. As it follows from Eq.(6), the state $|db_4\rangle$ coincides with the excited atomic bare state $|0\rangle$. Considering Eqs. (11-13) and taking into account that the contribution of the excited state into the atomic state-vector is described by the component $b_4^{(4)}$ of the dressed state vector $\vec{b}_k$ corresponding to the quasienergy $\lambda_4$, the relative contribution of the excited state $|db_4\rangle$ may be estimated as:

$$\left|\frac{b_4^{(4)}}{b_4^{(2)}}\right| \leq \left|\frac{\delta_{13}}{\sqrt{|W_1|^2+|W_2|^2}}\right| \tag{14}$$

As it follows from the obtained relation, the population of the excited state may be suppressed by increasing the amplitudes $W_1$ or/and $W_2$ of the Rabi frequencies of the laser pulses in Raman resonance. Obviously, the excited state also remains unpopulated when the system evolves in $\vec{b}_1 = |db_1\rangle$ dark state.

**3.1. Creation of coherent superposition states**

Let us now discuss creation of coherent superposition of ground states in the tripod-like atom without considerable excitation of the atom.

Considering the case of the positive Raman detuning along with the positive frequency chirp ($\beta > 0$), let us assume that the population of the atom is initially optically pumped into the one of the ground states, e.g. into the state $|1\rangle$. Taking into account the "dark-bright" superposition basis of Eqs.(6), the initial atomic state may be written as:

$$|\psi(-\infty)\rangle = |1\rangle = \frac{W_2|db_1\rangle + W_1^*|db_2\rangle}{\sqrt{|W_1|^2 + |W_2|^2}} \quad (15)$$

For this initial condition, we have for the coefficients $r_k(t) \cong r_k(-\infty)$ describing the statistical weights of the eigen-states in the dressed states representation of the atomic wave function in Eq.(8):

$$r_1 = \frac{W_2}{\sqrt{|W_1|^2 + |W_2|^2}} \quad r_2 = \frac{W_1^*}{\sqrt{|W_1|^2 + |W_2|^2}} \quad r_3 = 0 \quad r_4 = 0. \quad (16)$$

As it was discussed above, the dressed state $\vec{b}_1$, corresponding to the eigenvalue $\lambda_1 = 0$ coincides with the „dark" state $|db_1\rangle$ and does not change during the interaction. In the same time, as a result of the interaction, the dressed state $\vec{b}_4$ evolves from the „bright" state $|db_2\rangle$ to the state $|db_3\rangle$, with an additional $\pi$ phase-factor. The resulting state vector will have the following form in the „dark-bright" superposition basis and the bare states basis respectively:

$$|\psi(t\to+\infty)\rangle = \frac{1}{\sqrt{|W_1|^2+|W_2|^2}}(W_2|db_1\rangle - W_1^*|db_3\rangle)$$
$$= \frac{1}{\sqrt{|W_1|^2+|W_2|^2}}\left(\frac{|W_2|^2|1\rangle - W_2 W_1^*|2\rangle}{\sqrt{|W_1|^2+|W_2|^2}} - \frac{W_1^* W_3}{|W_3|}|3\rangle\right). \quad (17)$$

As it follows from the obtained equation, the final atomic wave function is a coherent superposition of the three ground states of the tripod-like atom. The contribution of different ground states into the obtained admixture is governed by the Rabi frequencies (intensities) of the two laser pulses in Raman resonance. Note that the excited state is absent in the final atomic wave function. During the interaction process there may be some temporary excitation of the atom, which however may be successfully suppressed by increasing the intensity of the pulses in Raman resonance, (see Eq.(14)).

A similar consideration may be provided for the case of negative Raman detuning and positive frequency chirp assuming that the tripod-like atom is prepared in the state $|3\rangle$ (see Fig.1(b)). In this case, the initial wave function of the atom may be written as:

$|\psi(-\infty)\rangle = |3\rangle = W_3^*/|W_3||db_3\rangle$, with coefficients $r_k(t) \cong r_k(-\infty)$: $r_1 = 0$; $r_2 = 0$; $r_3 = |W_3|/W_3$; $r_4 = 0$.

In this case, the dressed state $\vec{b}_4$ evolves from the state $|db_3\rangle$ to the state $|db_2\rangle$ with an additional $\pi$ phase-factor, so the final state vector becomes

$$|\psi(t\to+\infty)\rangle = -\frac{W_3^*}{|W_3|}|db_2\rangle = -\frac{W_3^*}{|W_3|\sqrt{|W_1|^2+|W_2|^2}}(W_1|1\rangle + W_2|2\rangle) \quad (18)$$

As it follows from Eq.(18), a coherent superposition of two ground states connected by the two laser pulses in Raman resonance is created. As in the previous case, the intensities of the pulses in Raman resonance govern the contribution of the two ground states in the coherent superposition. It is worth noting that if $W_1=W_2$, a maximum coherence of 0.5 is achieved in this scheme.

## 4. Results of Numerical Simulations

In the present Section, we analyze the considered above scheme of creation of coherent superposition states by numerical simulation of the system of Eqs.(5) for the density matrix elements. At this point, we take the duration of the pulses shorter than longitudinal and transverse relaxation times of the atomic system to confirm the conclusions based on the dressed state analysis of the previous Section.

The results of the numerical simulations are presented in Figs.(3 and 4) for the cases of positive and negative Raman detuning.

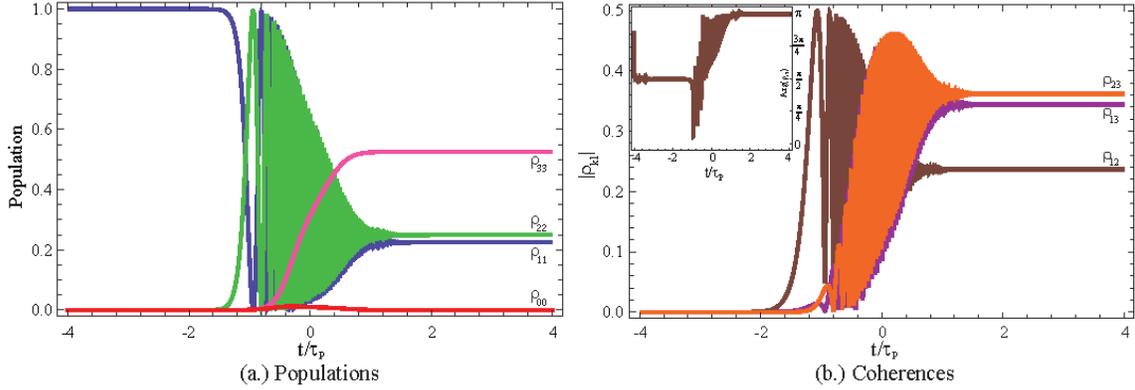

Fig.3. Time evolution of the populations and coherences in the case of positive Raman-detuning. The Inset: the absolute value of the coherence $\rho_{12}$. The parameters applied are: $W_1\tau_p=500$, $W_2\tau_p=475$, $W_3\tau_p=525$, $\beta\tau_p^2=2500$, $|\delta_{13}|\tau_p=250$.

As it follows from the numerical solutions, the dynamics of the states populations and coherences confirm the results of Section 3 based on the adiabatic consideration of the problem. Accordingly, in the case of the positive Raman detuning with initial preparation of the atom in the state $|1\rangle$, all the three ground states are populated at the end of the interaction: the population of the "bright" state is transferred into state $|3\rangle$, and the population of the "dark" superposition is left intact in the ground state (see Fig.3a). The time evolution of the atomic (bare) energy-eigenstate $|3\rangle$ corresponds to the dynamics of the single dressed state ($\vec{b}_4$). Consequently, no oscillations occurs in its time evolution. In contrary, the bare states $|1\rangle$ and $|2\rangle$ can be described as a superposition of the dressed states $\vec{b}_1$ and $\vec{b}_4$ with related *nonequal* eigenvalues $\lambda_1$ and $\lambda_4$. Due to this superposition of the dressed states, the time evolution of populations of the bare states displays an oscillatory character, (see Fig. 3a).

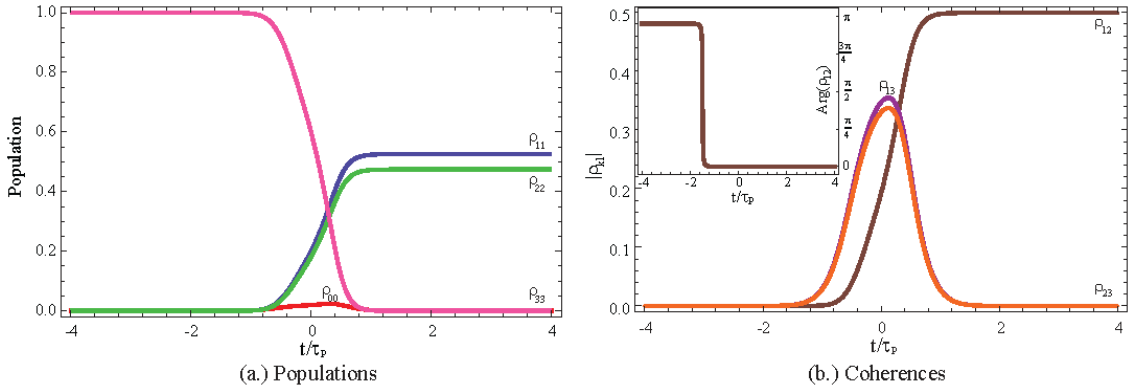

Fig. 4: Time evolution of the populations and coherences in the case of negative Raman-detuning. The Inset: the phase of the coherence $\rho_{12}$. The parameters used are the same as in Fig. 3.

In the case of negative Raman detuning, population of the initially populated state $|3\rangle$ is transferred into the "bright" superposition of the states $|1\rangle$ and $|2\rangle$. The time evolution of the populations in this case is connected to a single dressed state $\vec{b}_4$. As a result, no oscillations occur in the dynamics of the populations, (see Fig. 4a). It is important to note that, in both cases, the population of the excited state is negligable (and temporal), reaching merely 1-2% of the all atomic population.

We obtain the following expressions for the final values of the density matrix elements $\rho_{kl}^{fin} = \rho_{kl}(t \to +\infty), (k,l=0,...3)$ from Eqs.(17) and (18) corresponding to the cases of the positive ($\delta_{13} > 0$) and negative ($\delta_{13} < 0$) Raman detuning, respectively:

$$\delta_{13} > 0: \quad \rho_{00}^{fin} = 0; \; \rho_{11}^{fin} = \frac{|W_2|^4}{\left(|W_1|^2 + |W_2|^2\right)^2}; \; \rho_{22}^{fin} = \frac{|W_1|^2 |W_2|^2}{\left(|W_1|^2 + |W_2|^2\right)^2}; \; \rho_{33}^{fin} = \frac{|W_1|^2}{|W_1|^2 + |W_2|^2}$$

$$\rho_{12}^{fin} = \frac{|W_2|^2 W_1 W_2^*}{\left(|W_1|^2 + |W_2|^2\right)^2}; \; \rho_{13}^{fin} = \frac{|W_2|^2 W_1 W_3^*}{|W_3|^2 \left(|W_1|^2 + |W_2|^2\right)}; \; \rho_{23}^{fin} = \frac{|W_1|^2 W_2 W_3^*}{|W_3|^2 \left(|W_1|^2 + |W_2|^2\right)}; \quad (18)$$

and

$$\delta_{13} < 0: \quad \rho_{00}^{fin} = 0; \rho_{11}^{fin} = \frac{|W_1|^2}{|W_1|^2 + |W_2|^2}; \rho_{22}^{fin} = \frac{|W_2|^2}{|W_1|^2 + |W_2|^2}; \rho_{33}^{fin} = 0$$

$$\rho_{12}^{fin} = \frac{W_1 W_2^*}{|W_1|^2 + |W_2|^2}; \rho_{13}^{fin} = 0; \rho_{23}^{fin} = 0. \quad (19)$$

Dependence of the absolute value of the final induced coherence $\left|\rho_{12}^{fin}\right|$ on the ratio of the peak Rabi frequencies of the pulses in Raman resonance is presented in Fig.5 for both cases of positive and negative Raman detuning.

The results show an excellent agreement of the predictions based on the dressed states analysis (solid lines) with results of the numerical simulation of the Eqs(5) for the density matrix elements (the points).

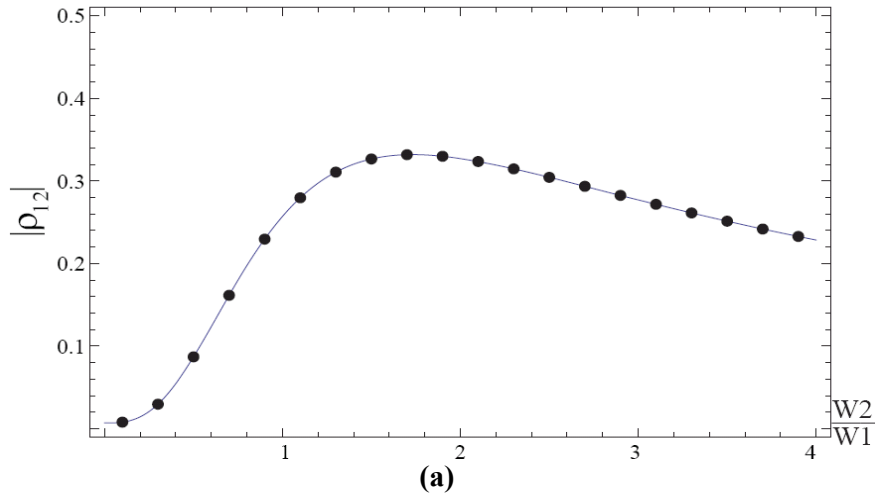

(a)

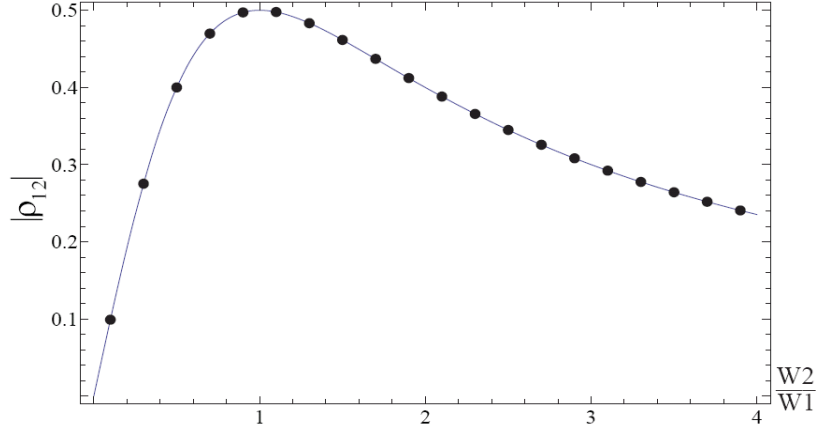

**(b)**

Fig.5. The resulting absolute values of the coherence $\rho_{12}$ versus the ratio of the peak values of the Rabi-frequencies $W_2/W_1$ for the cases of positive (a) and negative (b) Raman detuning calculated from the dressed states analysis. The dots are the results of the numerical solution of the master equation (5) in the absence of relaxation processes. The following values of the parameters are applied: $W_1\tau_p = 250$, $W_3\tau_p = 275$, $\beta\tau_p^2 = 2500$, $|\delta_{13}|\tau_p = 250$, $W_2\tau_p$ is varying.

## 5. Creation and control of the coherent superposition states in Doppler-broadened media

An inhomogeneous (e.g. Doppler broadening) of the atomic transition lines may limit the potential of schemes based on STIRAP to create coherent superposition states. We show at this point that using FC laser pulses in the proposed scheme allows creating and control of the coherent superposition states equally efficiently in both homogeneously and Doppler-broadened media.

A Doppler-broadened medium of a gas of tripod-structured atoms is modelled by averaging the created coherence between the ground states of the atoms over distribution of the resonance frequencies (velocities) of atoms in the gas at different values of temperature $T$ (for all three FC laser pulses propagating in a same direction). Considering a gas of $^{87}$Rb atoms at temperature $T$ and assuming Maxwell-Boltzmann distribution for the velocities of the atoms, we have for the probability distribution $P(\tilde{\Delta})$ for an atom to have single-photon detuning $\tilde{\Delta} = \Delta \tau_p$:

$$P(\tilde{\Delta}) = \sqrt{\frac{mc^2}{(2\pi)^3 kT (f_0 \tau_p)^2}} \exp\left[-\frac{mc^2 (\tilde{\Delta})^2}{8\pi^2 kT (f_0 \tau_p)^2}\right],$$

where $k$ is the Boltzmann constant, $m = 86.909\, u$ is the mass of $^{87}$Rb ($u$ being the atomic unit), $f_0 = 384.230\, THz$ is the frequency distance between the excited and the ground states ($F = 1$ and $F' = 0$ hyperfine states in the $D_2$ line of $^{87}$Rb), see Fig.6.

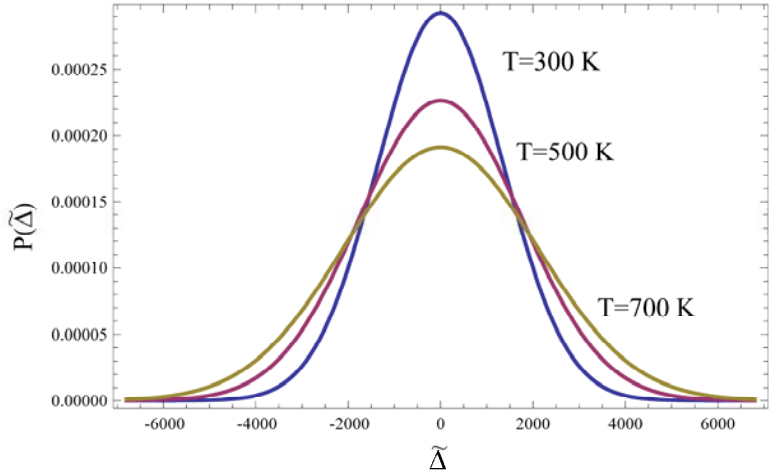

Fig. 6. The (normalized to unity) probability distribution for atomic gas at temperature equal to 300, 500 and 700 K.

The average values of the density matrix elements: populations and coherences $\langle \rho_{kl} \rangle$, (k,l = 0,1,2,3) are calculated numerically. First the master equation (5) is solved numerically for all values of $\tilde{\Delta}$ having nonzero probability in order to obtain the density matrix elements $\rho_{kl}^{final}(\tilde{\Delta})$ at the end of the interaction. The obtained values for populations and induced coherences are presented in Fig.7 as function of the single-photon detuning $\tilde{\Delta}$ presenting Doppler- shift of the transition lines due to the atomic motion.

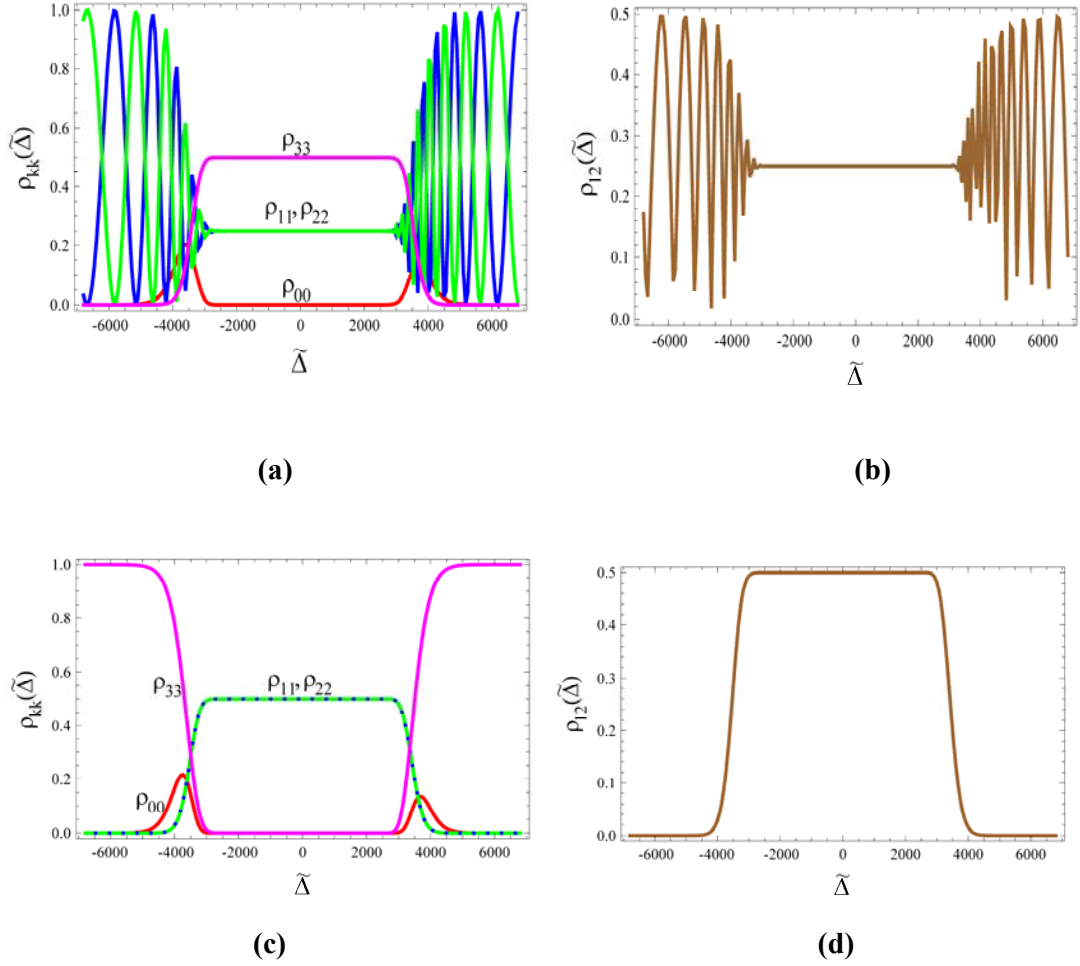

Fig.7. Final populations and coherences established in the atoms at the end of interaction with laser pulses as a function of the normalized single-photon detuning $\tilde{\Delta} = \Delta \tau_p$ for the cases of the positive: (a) and (b), and negative: (c) and (d) Raman detuning. The applied parameters are: $W_1 \tau_p = W_2 \tau_p = 250$, $W_3 \tau_p = 275$  $\beta \tau_p^2 = 1060$, $\delta_{13} \tau_p = 100$, $\tau_p = 10^{-6} s$.

As it is seen from Fig.7, there is a range of values of the Doppler shift $\tilde{\Delta}$ around zero, for which the final coherences (populations) do not depend on $\tilde{\Delta}$. This feature is due to the frequency-chirping of the laser pulses: as long as the Doppler shift is smaller than the frequency range $\left[-4\beta\tau_p, 4\beta\tau_p\right]$ covered by the chirp during the interaction (the interaction time may be approximated by $4\tau_p$), the velocity of the interacting atom does not have an impact on the resulting population and coherence distribution.

The averaging of the obtained solutions for the Doppler-broadened atomic gas is produced by numerically evaluating the integral

$$\langle \rho_{kl} \rangle = \int_{-\infty}^{\infty} P(\tilde{\Delta}) \rho_{kl}^{fin}(\tilde{\Delta}) d\tilde{\Delta}$$

The absolute value of the average induced coherence $|\langle \rho_{12} \rangle|$ established after the interaction with the laser field is presented in Fig.8 as a function of the normalized frequency span $\beta \tau_p^2$ of the laser pulses for different values of the gas temperature.

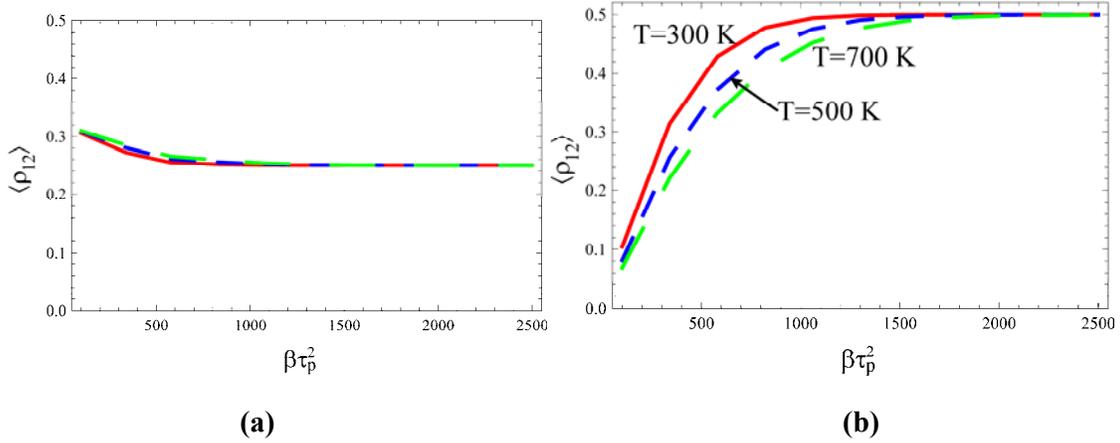

(a)          (b)

Fig.8. The average value of the final coherence between states $|1\rangle$ and $|2\rangle$ in the case of $\delta_{13} > 0$ (a), and $\delta_{13} < 0$ (b) as a function of the speed of chirp for the temperatures of the gas equal to 300 K(Red), 500 K (Blue) and 700 K (Green). The applied parameters are the same as in Fig.7.

As it can be seen from this Figure, the average value of the induced coherence does not depend on the Doppler-broadening for sufficiently large frequency span of the pulses during the interaction time due, for example to a sufficiently high speed of the frequency chirp.

## 6. Effect of the relaxation processes

In this Section, we discuss the influence of the relaxation processes on the efficiency of creation of the superposition states. At first step, we analyze the influence of the spontaneous decay of the excited state. The dependence of the states final populations and of the phase of the created coherence established after the interaction with the laser pulses is shown in Fig.9 as a function of the longitudinal relaxation rate.

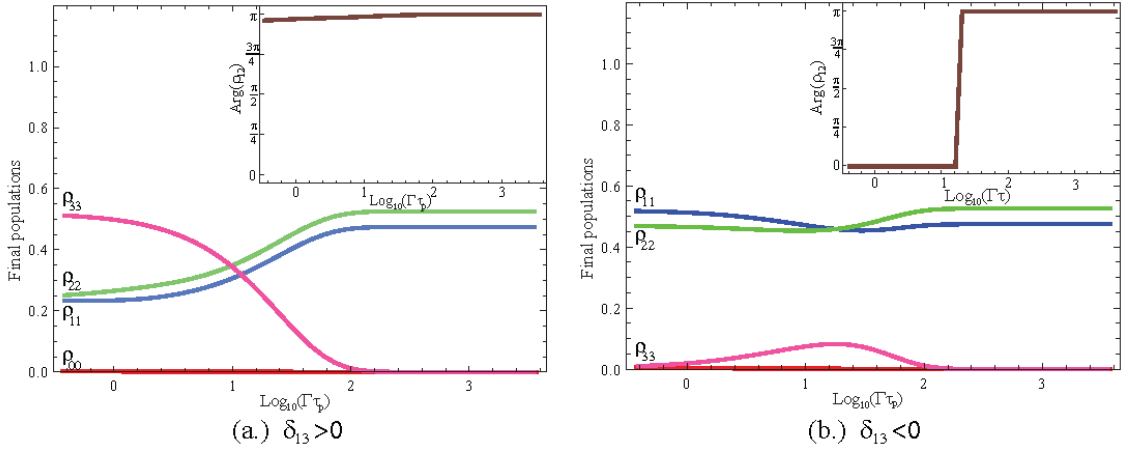

(a.) $\delta_{13}>0$    (b.) $\delta_{13}<0$

Fig.9. The final populations of the atomic states in the case of positive (a) and negative (b) Raman detuning, as a function of the dimensionless longitudinal relaxation rate in the absence of transverse relaxation processes. Insets: The phase of the created coherence $\rho_{12}$. The parameters used are: $W_1\tau_p=500$, $W_2\tau_p=475$, $W_3\tau_p=525$, $\beta\tau_p^2=2500$, $|\delta_{13}|\tau_p=250$.

One could anticipate a negligible influence of the spontaneous relaxation processes on the populations and coherences of the atom when no considerable excitation of the atom takes place. However, the results of the numerical simulations show, that even for the negligible excitation of the atom, the final populations and induced coherences depend on the longitudinal relaxation rate. The reason is linked to the optical coherences $\rho_{0k}$, ($k=1, 2, 3$) that are not negligibly small and depend on the longitudinal relaxation rate $\Gamma$, (see Eqs.(5)). As it can be seen from Fig.9, the larger the longitudinal relaxation rate (or longer the laser pulses), the more of the atomic poopulation is transferred (optically pumped) into states $|1\rangle$ and $|2\rangle$ connected by the laser pulses in Raman resonance. Insets in the Figures show behavior of the phase of the induced coherence, which imply that, for longer laser pulses, the atom ends up in the dark superposition in both cases of the positive and negative Raman detuning.

Since the quantum interference processes are the basis for the schemes of creation of coherent superposition states, the phase relations between the states probability amplitudes (the values and the phases of the corresponding coherencies) must play an important role in the considered processes. That is why a strong effect of the transverse relaxation (dephasing) processes may be anticipated on creation and control of the coherent superposition states as well as on the population transfer between the atomic states.

The final populations of the atomic states as a function of the transverse relaxation rate are presented in Fig.10 as a result of numerical simulation of the master equation, Eq.(5).

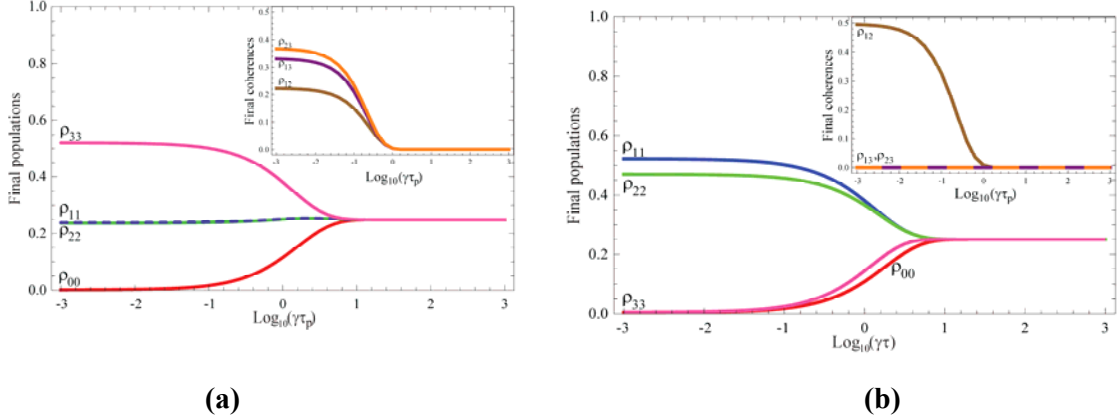

(a)          (b)

Fig. 10. The final populations of the atomic states in the case of positive (a) and negative (b) Raman detuning, as a function of the dimensionless transverse relaxation rate in the absence of longitudinal relaxation. Insets: The phase of the created coherence $\rho_{12}$. The parameters applied are: $W_1\tau_p=500$, $W_2\tau_p=475$, $W_3\tau_p=525$, $\beta\tau_p^2=2500$, $|\delta_{13}|\tau_p=250$.

It can be seen from the behavior of the populations that the effect of the transverse relaxation begins to be imperative at $\gamma\tau_p \approx 1/\sqrt{10}$. It means that the duration of the laser pulses $\tau_p$ has to be chosen much shorter than the transverse relaxation time scale $1/\gamma$: $\tau_p \leq 1/(\gamma\sqrt{10})$ to avoid the action of the dephasing on the creation of coherent superposition state. This is true for both cases of the positive and negative Raman detuning. If the interacting pulses are longer, the transverse relaxational processes destroy the adiabatic transfers. At large transverse relaxation rates, all the states are equally populated and there is no coherent superposition created as a result of the interaction (see Insets in Fig.10).

## 7. Robustness of the process

Let us now discuss robustness of the proposed scheme of creation of superposition states against variation of the main parameters of the laser radiation. As it follows from Fig.5 and Eqs.(18), (19), the absolute value of the created coherence $|\rho_{12}|$ depends on the ratio of the peak Rabi frequencies of the laser pulses in Raman resonance and does not depend on the phase relations between the laser pulses. One can control the induced coherence for example, by varying intensity of one of the laser pulses in Raman resonance leaving fixed the intensity of the second pulse in both cases of positive and negative Raman detuning. Note, that the induced coherence is robust against changes in the intensity of the laser pulse out of Raman resonance as soon as these changes do not violate the adiabaticity conditions, (see, for example [43,44]).

Robustness of the scheme against variations of the parameters of the pulses in Raman resonance may be provided utilizing laser pulses from a same source. In this case, at variation $\delta W$ of the peak Rabi frequences of the pulses, the variation of the ratio of the two peak Rabi frequencies may be estimated as $\delta(W_1/W_2) \approx W_1/W_2[1+\delta W(W_2-W_1)/(W_1 W_2)]$. As it follows from this relation, the dependence of the variation $\delta(W_1/W_2)$ on the parameter $\delta W$ is weak and hence, the robustness of the process is especially high in the case of close values of the peak Rabi frequencies of the pulses in Raman resonance: $W_1 \approx W_2$. In this case, the maximum value of the induced coherence $|\rho_{12}|$ equal to 0.5 is achieved for the negative Raman detuning (see Fig. 5(b)). Note that the considered schemes are also extremely robust against variations of the speed of the chirp, as usually takes place in the case of the adiabatic processes.

## 8. Conclusions

In conclusion, we have presented and investigated a scheme for creation of arbitrary coherent superposition of two and three ground states in a tripod-like atom without considerable excitation of the atom using FC laser pulses. The scheme is a generalization of the one proposed in Ref. [31] for coherent and adiabatic transfer of atomic population between the ground states in Λ-atom by a single frequency chirped laser pulse. In the present scheme, three similarly chirped laser pulses are applied to a tripod-like atom. Two of the pulses are in Raman resonance and the third one is out of Raman resonance providing transfer of the atomic populations between the ground states without considerable excitation of the atom. The created coherent superposition of the ground states is controlled by the ratio of the peak intensities of the laser pulses in Raman resonance.

We have analysed the applicability of the scheme in a Doppler-broadened medium of a gas composed of tripod-structured atoms by averaging the induced coherence over the velocity distribution of the atoms in the gas at different temperatures. The results show that the scheme is effective even for large widths of the Doppler – broadened transition lines if the frequency span of the laser pulses due to the chirp exceeds the width of the Doppler-broadening.

We have analyzed the influence of the relaxation processes on the population transfer and creation of the superposition states by numerical simulation of the master equation for the density matrix elements. We have shown that the considered scheme allows minimizing the effect of the spontaneous decay of the excited states by suppression of the population of the excited state. However, even under the condition of negligible excitation of the atom, for longer laser pulses, the longitudinal relaxation may influence the induced coherences and the resulting population distribution among the ground states. For laser pulses longer that the decay time of the excited state, optical pumping of the atom by a pair of the pulses in Raman resonance results in accumulation of the atomic population in dark superposition of the ground states linked by the laser pulses in Raman resonance.

As it may be anticipated, the transverse relaxation (dephasing) in the considered system have strong effect on the population transfer or creation of coherence in the tripod-structured atoms. The numerical simulation of the master equation has shown that already at duration of the laser pulses close to the dephasing constant of the medium the transverse relaxation processes destroy the adiabatic transfers. At larger transverse relaxation rates or longer laser pulses, all the states are equally populated as a result of interaction and there is no coherent superposition states created. The detrimental effect of the transverse relaxation may be avoided only by utilizing sufficiently short laser pulses.

We have shown that the proposed scheme is robust against small-to-medium variations in the intensities of the laser pulses. A possibility of robust creation of maximum coherence of 0.5 is demonstrated by using two laser pulses with close intensities from a common laser source.

The presented adiabatic scheme of coherence creation may find applications in the fields of quantum and nonlinear optics, atomic interferometry and in mapping and long-time storage of optical information in populations and coherences of metastable atomic states.

## Acknowledgements

The work was funded by the Research Fund of the Hungarian Academy of Sciences (OTKA) under contracts K 68240 and NN 78112.